\documentclass[aip,jmp, amsmath,amssymb,preprint]{revtex4-1}

\usepackage{graphicx}
\usepackage{dcolumn}
\usepackage{bm}

\begin{document}

\preprint{APS/}

\title{Excitation transport in quantum devices: analytical time-dependent non-equilibrium green function algorithm}

\author{I-Lin Ho}
\email{sunta.ho@msa.hinet.net}
\affiliation{ChiMei Visual Technology Corporation, Tainan 741, Taiwan, R.O.C.}

\date{\today}

\begin{abstract}
This research demonstrates analytical time-dependent non-equilibrium green function (TD-NEGF) algorithms to investigate dynamical functionalities of quantum devices, especially for photon-assisted transports.
Together with the lumped element model, we also study the effects of transiently-transferring charges to reflect the non-conservation of charges in open quantum systems,
and implement numerical calculations in hetero-junction systems composed of functional quantum devices and electrode-contacts (to the environment).
The results show that (\emph{i}) the current calculation by the analytical algorithms, versus those by conventional numerical integrals, presents superior numerical stability
on a large-time scale,
(\emph{ii}) the correction of charge transfer effects can better clarify non-physical transport issues, e.g. the blocking of AC signaling under the
assumption of constant device charges,
(\emph{iii}) the current in the long-time limit validly converges to the steady value obtained by standard time-independent density functional calculations,
and (\emph{iv}) the occurrence of the photon-assisted transport is well-identified.
\end{abstract}
\keywords{
 excitation transport, time-dependent non-equilibrium green function, photon, quantum dot}

\maketitle
\section{Introduction}
Photoelectric bioengineering - the use of photoelectric semiconductors as functional entities in biological systems - is heralded as an alternative option for signaling communications between organisms and physical devices in future biomedicines. In particular, research on quantum dots \cite{bio0,bio1} has already revealed a variety of biologically-oriented applications, e.g. drug discovery \cite{bio2,bio3}, disease detection \cite{bio4,bio5}, protein tracking \cite{bio6,bio7}, and intracellular reporting \cite{bio8,bio9}.
While a qualitative understanding of these complex processes has been accessed by perturbative electron-photon interactions associated with strong electron correlations \cite{qd_tb2},
the quantitative agreement between the first-principles theory and experiments is still unsatisfactory from the perspective of the ground-state density functional theory (DFT) \cite{qd_tb3,steady1}.

The majority of studies on quantum-dot electronics in recent years has focused on the time-dependent density functional theory (TDDFT) \cite{qd_tb4}, as it provides
a more rigorous theoretical foundation \cite{tddft1}. Its formalism may also be easily extended to cover the interaction of electrons with light or molecular environments
in open quantum systems via the time-dependent non-equilibrium green function (TDNEGF) technique \cite{thesis1,TDNEGF1},
e.g. for the photon-assisted transport and fluorescence of contacted atomic devices.
However, issues over numerical stability and the highly-demanding computational cost \cite{WBL2} make it difficult to apply the technique in mesoscopic biological systems.

To arrive at a computationally efficient but still predictive stage,
this research demonstrates analytical time-dependent non-equilibrium green function (TD-NEGF) algorithms for studying dynamical functionalities of quantum devices.
Together with introducing the analytical lumped element model \cite{orth1,QC1}, we also consider the effects of transiently-transferring charges.
Here, the lumped element model approximates a description of interactions of spatially-distributed transfer charges \cite{thesis1} into a capacitor-circuit topology,
significantly enhancing computation efficiency.

Numerical calculations are implemented in hetero-junction systems composed of functional quantum devices and electrode-contacts (to the environment), as indicated in Figure \ref{fig_TDNEGF}.
The central device is the Si-SiO$_{2}$ core-shell quantum dot,
where the core is designed in the strong confinement dimensions \cite{qd_tb4} (smaller than the Bohr radius; about 5 nm for silicon).
The silicon dioxide matrix is for the design of physical properties \cite{qd_tb2,qd_tb3}.
This work includes phosphorus impurities to enable low-voltage functionalities \cite{QD1}, and accounts for
the interactions between electrode-clusters and devices through properly defined self-energies.
For numerical treatments, we obtain the Kohn-Sham (KS) hamiltonian $\mathbf{h}$ and overlap matrices $\mathbf{s}$ of the ground-states for devices and Au electrodes
by standard time-independent density functional programs \cite{siesta1,siesta2}.
With the given $\mathbf{h}$ and $\mathbf{s}$, the transient properties of quantum transports are analyzed
using the present TD-NEGF algorithms \cite{TDNEGF1,TDNEGF2,codeF}.

This paper is organized as follows. Section \textrm{2} describes the theoretical algorithms.
Section \textrm{3} discusses the studies on numerical stability, transient-to-steady analyses, and photon-assisted transport.
Section \textrm{4} presents concluding remarks.
Appendix A describes the fundamental physical properties, from individual components to integrated device systems.
Appendix B calculates the conductance curve of the 4,4'-Bipyridine molecule with respect to photon energies,
and compares it with the Tien-Gordon approach,
for the purpose of identifying excitation transport dynamics.

\section{Time-dependent non-equilibrium green function for quasi-one-dimensional open quantum systems}

\begin{figure*}[ht]
\includegraphics[scale=0.375]{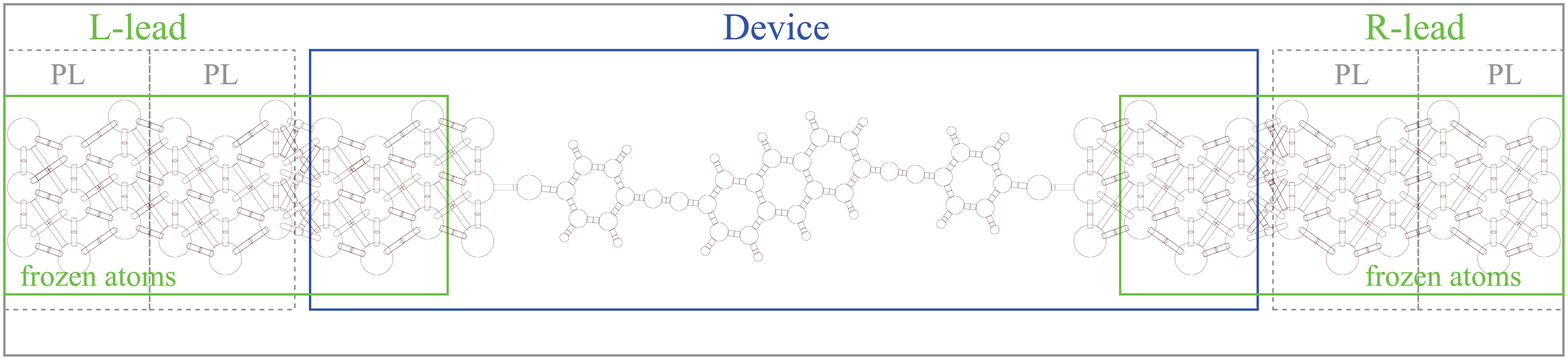}
\caption{ Schematic representation of general simulation setups for open quantum systems including side Au(111) electrodes and central quantum devices. }
\label{fig_TDNEGF}
\end{figure*}
Figure \ref{fig_TDNEGF} shows a regular open quantum system, including semi-infinite side electrodes and central quantum devices.
This system is partitioned by several electronically-functional areas, named as L-electrode (L), device (D), and R-electrode (R).
We describe the equation of motion (EOM) for electrons by the Heisenberg equation:
\begin{equation}
i\mathbf{\dot{\sigma}}\left( t\right) =\left[ \mathbf{h}\left( t\right) ,
\mathbf{\sigma }\left( t\right) \right]  \label{a1}
\end{equation}
where $\mathbf{h}\left( t\right) $ is the Kohn-Sham hamiltonian matrix, and
the square bracket on the right-hand side (RHS) denotes a commutator. The
matrix element of the single-electron density $\mathbf{\sigma }$ is defined
by $\sigma _{ij}\left( t\right) =\left\langle a_{j}^{\dagger
}(t)a_{i}(t)\right\rangle $, where $a_{j}^{\dagger }(t)$ and $a_{i}(t)$ are
the creation and annihilation operators for atomic orbitals $j$ and $i$ at
time $t$, respectively. On the basis of the atomic orbital sets for
electrons, the matrix representation of $\mathbf{\sigma }$ and $\mathbf{h}$
can be written as
\begin{equation}
\mathbf{h=}\left[
\begin{array}{cccc}
\mathbf{h}_{L} & \mathbf{h}_{LD} & 0 \\
\mathbf{h}_{DL} & \mathbf{h}_{D} & \mathbf{h}_{DR} \\
0 & \mathbf{h}_{RD} & \mathbf{h}_{R}
\end{array}
\right], \mathbf{\sigma =}\left[
\begin{array}{cccc}
 \mathbf{\sigma }_{L} & \mathbf{\sigma }_{LD} &
\mathbf{\sigma }_{LR} \\
 \mathbf{\sigma }_{DL} & \mathbf{\sigma }_{D} &
\mathbf{\sigma }_{DR} \\
 \mathbf{\sigma }_{RL} & \mathbf{\sigma }_{RD} &
\mathbf{\sigma }_{R}
\end{array}
\right]  \label{a2}
\end{equation}
We note that $\mathbf{m}_{L}$, $\mathbf{m}_{D}$, and $\mathbf{m}%
_{R}$ ($\mathbf{m}\in \left\{ \mathbf{h},\mathbf{\sigma }\right\} $)
represent the matrix blocks corresponding to left-electrode $L$, device $D$, and right-electrode $R$
partitions, respectively. Moreover, $\mathbf{h}_{LR}$ and $\mathbf{h}_{RL}$ are
ignored due to the distant separation between L and R electrodes in common
applications. It is noted that the holographic electron density
theorem and Runge-Gross theorem are applied for time-dependent electron dynamics \cite{TDNEGF1,TDNEGF2}, stating that the initial ground-state density
of the subsystem $\mathbf{\sigma }_{D}\left( t_{0}\right) $ can determine
all physical properties of systems at any time $t$. Hence, $\mathbf{h}$ and $\mathbf{\sigma }$ can be approximately expressed as
functions of $\mathbf{\sigma }_{D}\left( t\right) $ for a formally
closed-form equation of motion as described below.

Placing Eq. (\ref{a2}) into Eq. (\ref{a1}), we can write the equation of motion for $
\mathbf{\sigma }_{D}$ as
\begin{eqnarray}
i\dot{\sigma}_{D,mn} &=&\sum_{\ell \in D}\left( h_{D,m\ell }\sigma _{D,\ell
n}-\sigma _{D,m\ell }h_{D,\ell n}\right) -i\sum_{\alpha =L,R,N}Q_{\alpha ,mn}
\label{a3} \\
Q_{\alpha ,mn} &\equiv &i\sum_{k_{\alpha }\in \alpha }\left( h_{D\alpha
,mk_{\alpha }}\sigma _{\alpha D,k_{\alpha }n}-\sigma _{D\alpha ,mk_{\alpha
}}h_{\alpha D,k_{\alpha }n}\right)  \label{a4}
\end{eqnarray}
Here, $m$ and $n$ denote the atomic orbital in partition D, $k_{\alpha }$
denotes the state of $\alpha $ ($\alpha $=L, R) electrode, and $Q_{\alpha }$ is the
dissipation term due to the contacts of the device with electrodes L and R. The transient current through an electrode's
interfaces can be calculated by:
\begin{eqnarray}
I_{\alpha \in \left\{ L,R\right\} }\left( t\right) &=&-\int_{\alpha }d
\mathbf{r}\partial _{t}\rho \left( \mathbf{r},t\right) =-\sum_{k_{\alpha
}\in \alpha }\partial _{t}\sigma _{k_{\alpha }k_{\alpha }}\left( t\right)
\nonumber \\
&=&i\sum_{k_{\alpha }\in \alpha }\sum_{\ell \in D}\left( h_{D\alpha
,k_{\alpha }\ell }\sigma _{\alpha D,\ell k_{\alpha }}-\sigma _{D\alpha
,k_{\alpha }\ell }h_{\alpha D,\ell k_{\alpha }}\right)  \nonumber \\
&=&-tr\left[ Q_{\alpha }(t)\right]  \label{a5}
\end{eqnarray}

\subsection{Expressions of the dissipation function $Q_{\alpha}$ using the Green function formalism}
To calculate the dissipation term $Q_{\alpha }$ in EOM and the transient current
equation, this work uses the time-dependent non-equilibrium Green function (TDNEGF) formalism.
It is noted that a replacement for the overlap-matrix by the identity matrix is proceeded
by redefining the device's hamiltonian \cite{book1}(Ch. 8.1.2): $\mathbf{h}_{D}-E\mathbf{s}_{D}=\mathbf{h}_{D}-E(\mathbf{s}_{D}-\mathbf{I})-E%
\mathbf{I}=\mathbf{h}_{D}^{\prime }-E\mathbf{I}$.
The expression of the dissipation function $Q_{\alpha }$ hence can be derived as \cite{TDNEGF1}:
\begin{equation}
Q_{\alpha ,mn}(t)=-\sum_{\ell \in D}\int_{-\infty }^{\infty }d\tau \left[
G_{D,m\ell }^{<}\left( t,\tau \right) \Sigma _{\alpha ,\ell n}^{A}\left(
\tau ,t\right) +G_{D,m\ell }^{R}\left( t,\tau \right) \Sigma _{\alpha ,\ell
n}^{<}\left( \tau ,t\right) +H.c.\right]  \label{a7}
\end{equation}
where the lesser Green functions $\mathbf{G}
^{<}$ and the retarded Green functions $\mathbf{G}^{R}$ in Eq. (\ref{a7}) are determined via Kadanoff-Baym equations
\cite{TDNEGF1,kb1}:
\begin{eqnarray}
i\frac{d}{dt}\mathbf{G}_{D}^{R}\left( t,t^{\prime }\right) &=&\delta \left(
t-t^{\prime }\right) +\mathbf{h}_{D}\left( t\right) \mathbf{G}_{D}^{R}+
\mathbf{\Sigma }^{R}\cdot \mathbf{G}_{D}^{R}  \label{a8} \\
i\frac{d}{dt}\mathbf{G}_{D}^{<}\left( t,t^{\prime }\right) &=&\left[ \mathbf{
\Sigma }^{R}\cdot \mathbf{G}^{<}+\mathbf{\Sigma }^{<}\cdot \mathbf{G}^{A}
\right] \left( t,t^{\prime }\right) +\mathbf{h}\left( t\right) \mathbf{G}
^{<}\left( t,t^{\prime }\right)  \label{a9}
\end{eqnarray}
with notations $\left[ f\cdot g\right] \left( t,t^{\prime
}\right) =\int_{t_{0}}^{\infty }d\bar{t}f(t,\bar{t})g(\bar{t},t^{\prime })$
, $\mathbf{\Sigma }^{\lessgtr ,A,R}=\sum_{\alpha }\mathbf{\Sigma }_{\alpha
}^{\lessgtr ,A,R}$, and $f^{A}\left( t,t^{\prime }\right) =
\left[ f^{R}\left( t^{\prime },t\right) \right] ^{\dagger }$ \ $\left( f\in
G,\Sigma \right) $. The advanced self-energy $\mathbf{\Sigma }_{\alpha }^{A}$
and the lesser self-energy $\mathbf{\Sigma }_{\alpha }^{<}$ for electrode $\alpha$ by definition are:
\begin{eqnarray}
\mathbf{\Sigma }_{\alpha }^{A}\left( t,t^{\prime }\right) &=&i\Theta \left(
t^{\prime }-t\right) \mathbf{h}_{D\alpha }(t)\exp \left\{
i\int_{t}^{t^{\prime }}\mathbf{h}_{\alpha }\left( \bar{t}\right) d\bar{t}
\right\} \mathbf{h}_{\alpha D}(t^{\prime })  \label{a10} \\
\mathbf{\Sigma }_{\alpha }^{<}\left( t,t^{\prime }\right) &=&i\mathbf{h}
_{D\alpha }(t)f_{\alpha }\left( \mathbf{h}_{\alpha ,t=t_{0}}\right) \exp
\left\{ i\int_{t}^{t^{\prime }}\mathbf{h}_{\alpha }\left( \bar{t}\right) d
\bar{t}\right\} \mathbf{h}_{\alpha D}(t^{\prime })  \label{a11}
\end{eqnarray}
Here, $\Theta \left( t^{\prime }-t\right) $ is the Heaviside step function, $
\mathbf{h}_{\alpha }$ is the Kohn-Sham matrix of the isolated electrode $\alpha$, and $
f_{\alpha }$ is the Fermi distribution function for $\alpha \in L,R$.

\subsection{Wide-band limit approximation for the dissipation function $\mathbf{Q}_{\alpha}$}

For efficient computations of the equation of motion in Eqs. (\ref{a8}) and (\ref
{a9}), we introduce the wide-band limit (WBL) approximation \cite{WBL1}
for L and R electrodes under conditions \cite{TDNEGF1,WBL2}: (1) the bandwidths of the electrodes are
larger than the coupling strength between the device and L or R electrode; (2) the broadening matrix (the imaginary part of
self-energy, as defined below) is assumed to be energy-independent, resulting
in the requirement for an electrode's density of state and device-electrode
couplings to be slowly varying in energy; and (3) the level shifts of
electrodes via bias are approximated to be constant for all energy levels.

Through the conditions for the wide-band limit approximation, the self-energy
is split up into two real matrices: one is the hermitian matrix $\mathbf{\Lambda}
_{\alpha }$ representing level shift, and the other is the anti-hermitian matrix $\mathbf{\Gamma} _{\alpha }$ representing level broadening. Specifically, Eqs. (\ref
{a10}) and (\ref{a11}) are:
\begin{equation}
\mathbf{\Sigma }_{\alpha }^{R,A}\left( t,t^{\prime }\right) =\left( \mathbf{
\Lambda }_{\alpha }\mp i\mathbf{\Gamma }_{\alpha }\right) \delta \left(
t-t^{\prime }\right)   \label{a14}
\end{equation}
where $\mathbf{\Lambda }_{\alpha }$ and $\mathbf{\Gamma }_{\alpha }$ obey the Kramers-Kronig relation \cite{kk1}.
The dissipation term for electrodes L and R now is \cite{thesis1}:
\begin{equation}
\mathbf{Q}_{\alpha }(t)=\mathbf{K}_{\alpha }(t)+\mathbf{K}_{\alpha
}^{\dagger }(t)+\left\{ \mathbf{\Gamma }_{\alpha },\mathbf{\sigma} \left( t\right)
\right\} +i\left[ \mathbf{\Lambda }_{\alpha },\mathbf{\sigma} \left( t\right) \right]
\label{a15}
\end{equation}
with the definition of $\mathbf{K}_{\alpha }(t)$ as:
\begin{eqnarray}
&&\mathbf{K}_{\alpha }(t)=-\frac{2i}{\pi }\mathbf{U}_{\alpha
}(t)\int_{-\infty }^{\infty }\frac{f_{\alpha }\left( \epsilon \right)
e^{i\epsilon t}}{\epsilon -\mathbf{h}_{D}(0)-\sum_{\alpha '}\left( \mathbf{%
\Lambda }_{\alpha '}-i\mathbf{\Gamma }_{\alpha '}\right) }d\epsilon \mathbf{%
\Gamma }_{\alpha }  \label{a16} \\
&&-\frac{2i}{\pi }\int_{-\infty }^{\infty }\left[ \mathbf{I-U}_{\alpha
}(t)e^{i\epsilon t}\right] \frac{f_{\alpha }\left( \epsilon \right) }{%
\epsilon -\mathbf{h}_{D}(t)-\sum_{\alpha '}\left( \mathbf{\Lambda }_{\alpha
'}-i\mathbf{\Gamma }_{\alpha '}\right) +V_{\alpha }\left( t\right) \mathbf{I}}%
d\epsilon \mathbf{\Gamma }_{\alpha }  \nonumber
\end{eqnarray}
and
\begin{equation}
\mathbf{U}_{\alpha }(t)=e^{-i\int_{0}^{t}\left[ \mathbf{h}_{D}(\bar{t}
)+\sum_{\alpha '}\left( \mathbf{\Lambda }_{\alpha '}-i\mathbf{\Gamma }_{\alpha
'}\right) -V_{\alpha }\left( \bar{t}\right) \mathbf{I}\right] d\bar{t}}
\label{a16w}
\end{equation}
Together with EOM for $\mathbf{\sigma }_{D}(t)$ in Eqs. (\ref{a3}) and (\ref{a4}), one is prepared to calculate the transient electron density of the device and the boundary currents in Eq. (\ref{a5}).

\subsection{Calculations of self-energy matrices $\mathbf{\Lambda}$ and $\mathbf{\Gamma}$ in $\mathbf{Q}_{\alpha}$}
In principle, we can formulate the retarded self-energy for contact with electrode $\alpha$ in the energy domain \cite{thesis1} as:
\begin{equation}
\mathbf{\Sigma }_{\alpha }^{r}(E)=\mathbf{h}_{D\alpha } \mathbf{G}^{r}_{\alpha}(E) \mathbf{h}_{\alpha D}
\label{selfer}
\end{equation}
Considering the semi-infinite electrodes, the periodic Au(111) lattices can be divided into principle layers (PLs) along the transport direction (see Fig \ref{fig_TDNEGF}).
Here, we choose PLs to be wide enough so that only interactions between the nearest PLs need to be considered;
i.e. the coupling matrix $\mathbf{h}_{D\alpha}$ between contact $\alpha$ and device region $D$ will be restricted to one PL.
Consequently only the surface block of $\mathbf{G}^{r}_{\alpha}$, i.e. the surface green function $\mathbf{G}^{r,s}_{\alpha}$, is needed for calculating Eq. (\ref{selfer}).
This work adopts an iterative method \cite{surfG} to calculate the surface green function that includes properties of the semi-infinite lattices.
Specifically, we calculate the self-energy matrices $\mathbf{\Gamma}$ and $\mathbf{\Lambda}$ for wide-band approximation at the Fermi level as
\begin{equation}
\mathbf{h}_{D\alpha } \mathbf{G}^{r,s}_{\alpha}(E_{F}) \mathbf{h}_{\alpha D}=\mathbf{\Lambda}_{\alpha}-i\mathbf{\Gamma}_{\alpha}
\label{selfer2}
\end{equation}

\subsection{Analytical formulae of the $\mathbf{K}_{\alpha }$ term in $\mathbf{Q}_{\alpha}$}
On the calculation of the function $\mathbf{K}_{\alpha }$ in Eq. (\ref{a16}%
), this work introduces two approximations for enhancing numerical
stability, accuracy, and efficiency in large(-time-space)-scale simulations:
\begin{eqnarray}
f_{\alpha }\left( \epsilon \right) &\doteqdot &%
\begin{array}{ccc}
\lceil & 1 & ,\epsilon \leq \mu _{\alpha }-k_{b}T \\
\lfloor & \exp \left( -\frac{\epsilon -\mu _{\alpha }+k_{b}T}{k_{b}T}\right)
& ,\epsilon >\mu _{\alpha }-k_{b}T%
\end{array}
\label{add1} \\
\Gamma \left( 0,z\right) &\doteqdot &%
\begin{array}{ccc}
\lceil & e^{-z}\sum_{n=0}^{\infty }\frac{(-1)^{n}n!}{z^{n+1}} & ,\left\vert
z\right\vert \gg 1 \\
\lfloor & \Gamma \left( 0,z\right) & ,else%
\end{array}
\label{add2}
\end{eqnarray}%
Here, $\Gamma \left( n,z\right) $ is the incomplete gamma function, and $\mu
_{\alpha }$ is the total chemical potential for the Fermi distribution
function $f_{\alpha }$ of the electrode $\alpha $. For simplicity, the
variable $\mathbf{h}_{\alpha,eff}(t)\equiv \mathbf{h}_{D}(t)+\sum_{\alpha '}\left(
\mathbf{\Lambda }_{\alpha '}-i\mathbf{\Gamma }_{\alpha '}\right) -V_{\alpha
}\left( t\right) \mathbf{I}$ is introduced to define the effective
hamiltonian. By expressing $\mathbf{h}_{\alpha
,eff}(t)=\mathbf{\phi }_{\alpha ,t}\mathbf{\kappa }_{\alpha }(t)\mathbf{\phi
}_{\alpha ,t}^{-1}$ with its eigenvector matrix $\mathbf{\phi }_{\alpha ,t}$
and the diagonal eigenvalue matrix $\mathbf{\kappa }_{\alpha }(t)$, we can analytically rewrite
equation (\ref{a16}):

\begin{eqnarray}
\mathbf{K}_{\alpha }(t) &=&-\frac{2i}{\pi }\mathbf{U}_{\alpha }(t)\mathbf{%
\phi }_{\alpha ,0}\int_{-\infty }^{\infty }f_{\alpha }\left( \epsilon
\right) e^{i\epsilon t}\left[ \epsilon \mathbf{I}-\mathbf{\kappa }_{\alpha
}(0)\right] ^{-1}d\epsilon \mathbf{\phi }_{\alpha ,0}^{-1}\mathbf{\Gamma }%
_{\alpha }  \label{add3} \\
&&+\frac{2i}{\pi }\mathbf{U}_{\alpha }(t)\mathbf{\phi }_{\alpha
,t}\int_{-\infty }^{\infty }f_{\alpha }\left( \epsilon \right) e^{i\epsilon
t}\left[ \epsilon \mathbf{I}-\mathbf{\kappa }_{\alpha }(t)\right]
^{-1}d\epsilon \mathbf{\phi }_{\alpha ,t}^{-1}\mathbf{\Gamma }_{\alpha }
\nonumber \\
&&-\frac{2i}{\pi }\mathbf{\phi }_{\alpha ,t}\int_{-\infty }^{\infty
}f_{\alpha }\left( \epsilon \right) \left[ \epsilon \mathbf{I}-\mathbf{%
\kappa }_{\alpha }(t)\right] ^{-1}d\epsilon \mathbf{\phi }_{\alpha ,t}^{-1}%
\mathbf{\Gamma }_{\alpha }  \nonumber \\
&\equiv &-\frac{2i}{\pi }\left[ \mathbf{U}_{\alpha }(t)\mathbf{\phi }%
_{\alpha ,0}\mathbf{\Theta }_{\alpha 1}\mathbf{\phi }_{\alpha ,0}^{-1}-%
\mathbf{U}_{\alpha }(t)\mathbf{\phi }_{\alpha ,t}\mathbf{\Theta }_{\alpha 2}%
\mathbf{\phi }_{\alpha ,t}^{-1}+\mathbf{\phi }_{\alpha ,t}\mathbf{\Theta }%
_{\alpha 3}\mathbf{\phi }_{\alpha ,t}^{-1}\right] \mathbf{\Gamma }_{\alpha }
\nonumber
\end{eqnarray}%
The elements of the diagonal matrices $\mathbf{\Theta }_{\alpha 1}$, $%
\mathbf{\Theta }_{\alpha 2}$, and $\mathbf{\Theta }_{\alpha 3}$ can be analytically calculated by:%
\begin{equation}
\Theta _{\alpha 1,ii}(t)=\Theta _{\alpha 2,ii}(t)|_{\kappa _{\alpha
,ii}(t)\rightarrow \kappa _{\alpha ,ii}(0)}  \label{add4}
\end{equation}%
\begin{eqnarray}
\Theta _{\alpha 2,ii}(t) &=&e^{i\kappa _{\alpha ,ii}\cdot t}\left[ \Gamma %
\left[ 0,-i\left( \epsilon _{L}-\kappa _{\alpha ,ii}\right) t\right] -\Gamma %
\left[ 0,-i\left( \epsilon _{M}-\kappa _{\alpha ,ii}\right) t\right] \right]
\label{add5} \\
&&+e^{i\kappa _{\alpha ,ii}\cdot t}[ln\left( \epsilon _{M}-\kappa _{\alpha
,ii}\right) -ln\left( \epsilon _{L}-\kappa _{\alpha ,ii}\right)   \nonumber
\\
&&-ln\left( it\left( \kappa _{\alpha ,ii}-\epsilon _{M}\right) \right)
+ln\left( it\left( \kappa _{\alpha ,ii}-\epsilon _{L}\right) \right) ]
\nonumber \\
&&+e^{i\kappa _{\alpha ,ii}\cdot t}e^{\frac{\epsilon _{M}-\kappa _{\alpha
,ii}}{k_{b}T}}\left[ \Gamma \left[ 0,\left( \kappa _{\alpha ,ii}-\epsilon
_{M}\right) \left( it-\beta \right) \right] -\Gamma \left[ 0,\left( \kappa
_{\alpha ,ii}-\epsilon _{H}\right) \left( it-\beta \right) \right] \right]
\nonumber \\
&&+e^{i\kappa _{\alpha ,ii}\cdot t}e^{\frac{\epsilon _{M}-\kappa _{\alpha
,ii}}{k_{b}T}}[ln\left( \epsilon _{H}-\kappa _{\alpha ,ii}\right) -ln\left(
\epsilon _{M}-\kappa _{\alpha ,ii}\right)   \nonumber \\
&&-ln\left( \left( it-\beta \right) \left( \kappa _{\alpha ,ii}-\epsilon
_{H}\right) \right) +ln\left( \left( it-\beta \right) \left( \kappa _{\alpha
,ii}-\epsilon _{M}\right) \right) ]  \nonumber
\end{eqnarray}%
\begin{eqnarray}
\Theta \alpha _{3,ii}(t) &=&e^{\frac{\epsilon _{M}-\kappa _{\alpha ,ii}(t)}{%
k_{b}T}}\Gamma \left[ 0,\frac{\epsilon _{M}-\kappa _{\alpha ,ii}(t)}{k_{b}T}%
\right] -e^{\frac{\epsilon _{M}-\kappa _{\alpha ,ii}(t)}{k_{b}T}}\Gamma %
\left[ 0,\frac{\epsilon _{H}-\kappa _{\alpha ,ii}(t)}{k_{b}T}\right]
\nonumber \\
&&+ln\left[ \epsilon _{M}-\kappa _{\alpha ,ii}(t)\right] -ln\left[ \epsilon
_{L}-\kappa _{\alpha ,ii}(t)\right]   \label{add6}
\end{eqnarray}%
where the energies $\epsilon _{L}$ and $\epsilon _{H}$ are the lower integral boundary
and the higher integral boundary, respectively. $\epsilon _{M}=\mu _{\alpha
}-k_{b}T$ is the condition boundary of the approximation function in Eq. (%
\ref{add1}), and $\beta =k_{b}^{-1}T^{-1}$ is the inverse temperature. The
complex natural logarithm of $z$ denotes $ln\left( z\right) =ln\left(
\left\vert z\right\vert \right) +i\cdot arg\left( z\right) $.

The wide-band dissipation function $\mathbf{Q}_{\alpha }$ in Eq. (\ref{a15})
now can be efficiently calculated with the given device hamiltonian, the
device reduced density matrix, the self-energies containing the effect of
the leads, and the analytical $\mathbf{K}_{\alpha }$ formulae.

\subsection{Correction of the device Hamiltonian for transient variations of electron densities
using the lumped element model}
To consider the effects of transiently-transferring charges $\delta q$ in the open quantum system,
the device hamiltonian can be expressed in the perturbative form \cite{hform1} of:
\begin{equation}
\mathbf{h}_{D}=\mathbf{h}_{D}^{0}\left( q_{0}\right) +\delta \mathbf{h}%
_{D}\left( \delta q\right)  \label{hd_t1}
\end{equation}%
Here, the change of electron density can be computed via the density matrix
$\mathbf{\sigma }_{D}$ in Eq. (\ref{a3})
\begin{equation}
\delta n\left( \vec{r}\right) =\sum_{\mu \nu }Re[\rho _{\mu \nu }\chi _{\mu
}\left( \vec{r}\right) \chi _{\nu }^{\ast }\left( \vec{r}\right)
]-n_{0}\left( \vec{r}\right)  \label{mullq1}
\end{equation}%
as a function of spatial variable $\vec{r}$, or, alternatively, by
\begin{equation}
\delta q_{i}=\sum_{\mu \in \{i\}}\sum_{\nu }Re[\rho _{\mu \nu }s_{D,\nu \mu
}]-q_{0,i}  \label{mullq2}
\end{equation}%
using the atom-site notations. Here, $n_{0}\left( \vec{r}\right) $ and $q_{0,i}$ are the reference
charges chosen for neutrality, $\mathbf{s}_{D}$ is the device overlap
matrix, and $\chi _{i}\left( \vec{r}\right) $ is a set of local basis
functions used in the tight-binding formulation. According to the Taylor expansion of the total energy around the reference density, this change of charge density
can result in corrections to the Hartree and the exchange-correlation potentials
\cite{perturb1,perturb2} for the device hamiltonian as in Eq. (\ref{hd_t1}),
and it is continuously renewed with the transient density matrix in Eq. (\ref{a3}). Herein, we simplify
the correction of the device hamiltonian $\delta \mathbf{h}_{D}$ by
retaining only the Hartree potential $\delta V_{H}$ (assuming the
exchange-correlation term is insignificant in the mean-field criterion), which obeys the three-dimensional Poisson
equation
\begin{equation}
\nabla ^{2}\delta V_{H}(r)=-\delta n\left( \vec{r}\right)  \label{Poisson}
\end{equation}%
with the boundary conditions imposed by the lead potentials. Since the conventional
Poisson solution is based on spatially-discretized grids ($> N^{3}$ grids for $N$-atom systems) with numerically iterative processes,
the computations can be significantly time-consuming for large systems. Thus, it is convenient to study another
efficient analytical model.

On the basis of the success of the muffin-tin (MT) approximation, the total excess charge $\delta q_{i}$ is assumed to collectively locate
within a spherical region (MT-sphere) surrounding its nucleus $i$. The interactive charges inside different MT-spheres are considered as
capacitance effects \cite{datta1}. All MT-spheres ($N$-variables) of the system construct a capacitance-circuit
architecture in the lumped element model that supplies an analytical solution for the Poisson
equation \cite{blockbook1}. In principle,
the capacitances are treated as a
combination of the electrostatic capacitance $c_{e}$ and quantum capacitance
$c_{Q}$ \cite{datta1}. Herein, we assume the quantum capacitance to be less
dominant than the electrostatic capacitance for $\delta q_{i}$ and
ignore it in our work.

Replacing the spatial solution ($\nabla _{r}^{2}$) of the Poisson equation by
the atom-site notations ($\nabla _{i}^{2}$) for the lumped element model
\cite{orth1,orth2}, we can rewrite Eq. (\ref{Poisson}) by a matrix-form
equation $\mathbf{\hat{C}\vec{V}}=\mathbf{\vec{Q}}$
\begin{eqnarray}
c_{ij} &=&4\pi \epsilon \frac{\bar{a}_{ij}^{2}}{|r_{ij}|}\left( 1+\frac{%
\bar{a}_{ij}^{2}}{|r_{ij}|^{2}-2\bar{a}_{ij}^{2}}+...\right)
\label{circuit1} \\
\hat{C}_{ij} &=&\sum_{k\in \{1NN\}_{i,con}}\delta _{i,j}c_{ik}+\sum_{k\in \{1NN\}_{i}}\delta _{i,j}c_{ik}-\delta
_{j,k}c_{ij}  \label{circuit2} \\
\vec{Q}_{i} &=&e\cdot \delta q_{i}+e\cdot \delta q_{d,i}+\sum_{k\in
\{1NN\}_{i,con.}}\delta _{j,k}c_{ij}V_{con.j}  \label{circuit3}
\end{eqnarray}%
Here, the matrix elements of $\mathbf{\hat{C}}$ are calculated in a two-center approximation as proposed in the tight-binding approach \cite{thesis1}, obeying the formal condition
$e\delta q_{i}+e\delta q_{d,i}=\sum_{j}c_{ij}(\delta
V_{i}-\delta V_{j})$.
The notation $\{1NN\}_{i}$ is the group of the first nearest-neighbor (NN) atoms in the
device region for atom $i$, and $\{1NN\}_{i,con.}$ is the group of the first
nearest-neighbor atoms in the lead region. Herein, $c_{ij}$ defines
the capacitance between two ideal metal spheres, $|r_{ij}|$ is the spatial distance between atoms $i$ and $j$, and
$\bar{a}_{ij}$ is the effective muffin-tin radius for atoms $i$ and $j$ and is defined by $\bar{a}_{ij}=(r_{MT,i}+r_{MT,j})/4$ in this work.
Moreover, $\mathbf{\vec{V}}\equiv (\delta V_{1},\delta V_{2},...,\delta V_{N})$ is the
potential vector with the components being deviations of electrostatic potentials on atom-sites
$i\in \left\{ 1,...,N\right\} $. $V_{con.j}$ is the potential of lead atom $j
$ imposed by boundary conditions. $\delta q_{i}$ is the variation of the charge density
obtained by Eq. (\ref{mullq2}), and $\delta q_{d,i}$ represents the defect
charge for atom $i$. By linear algebra the potential vector $\mathbf{\vec{V}}$ can be easily
solved using $\mathbf{\vec{V}}=\mathbf{\hat{C}}^{-1}\mathbf{%
\vec{Q}}$.
For instance, in a 1-dimensional homogeneous system having 4
atoms L-A-A-R, the capacitance between nearby atoms is denoted as $c$, and the biases are denoted
as $v_{L}$ and $v_{R}$ for lead atoms L and R, respectively. There are no excess
charges ($\delta q=0$) inside the MT-sphere of device atoms A. In this way, the 2x2
capacitance matrix has components $\hat{C}_{11}=\hat{C}_{22}=2c$ and $\hat{C}%
_{12}=\hat{C}_{21}=-c$, and the charge vector is $\mathbf{\vec{Q}}^{t}=[
\begin{array}{cc}
cv_{L} & cv_{R}%
\end{array}] $. One then can obtain the electrostatic potentials for the two
atoms A as $\mathbf{\vec{V}}^{t}=[
\begin{array}{cc}
2v_{L}+v_{R} & v_{L}+2v_{R}%
\end{array}] /3$, which agree with the free-space Poisson solution.

\begin{figure}[tbp]
\begin{center}
\includegraphics[scale=0.55]{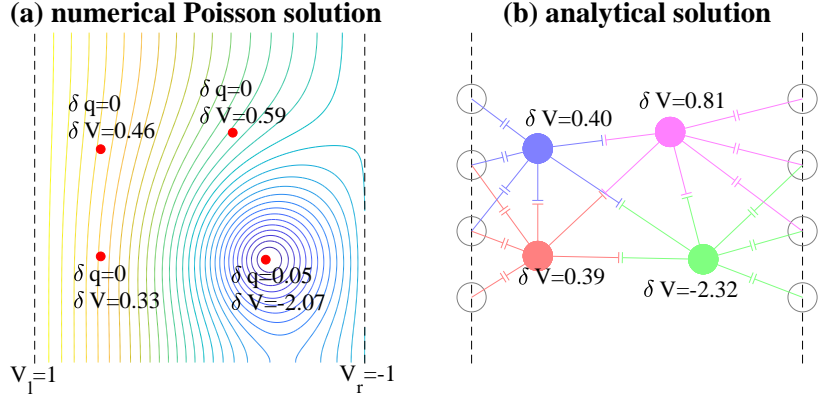}
\end{center}
\caption{Profile of Hartree potential $\delta V_{i}$ for the structure with area $9\times 9$-$\AA^{2}$, solved by
(a) the numerical Poisson solution and (b) the analytical solution. Four atoms with specified charges $\delta q$
 are placed between two leads and have $r_{MT}=1\AA$. Plot (a) illustrates the spatial distribution function $\delta V(r)$ by the contour curves.
 Plot (b) depicts the solution of the lumped element $\delta V_{i}$, where the boundary condition of the electrostatic potential is represented by four lead atoms.  } \label{poissonsolve1}
\end{figure}

\begin{figure}[tbp]
\begin{center}
\includegraphics[scale=0.55]{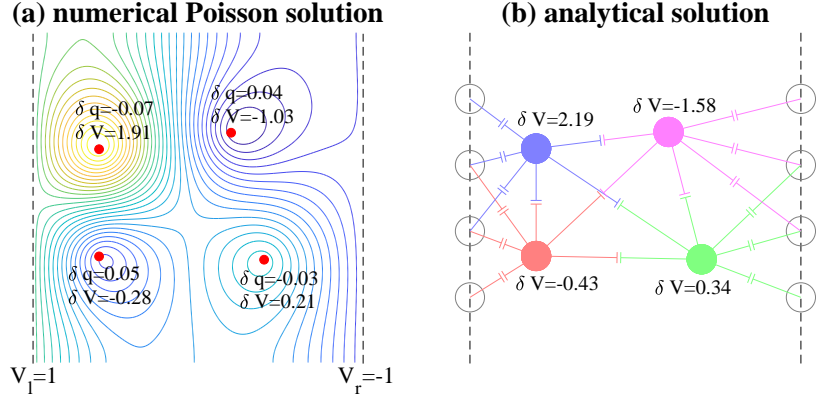}
\end{center}
\caption{Profile of Hartree potential $\delta V_{i}$ for the structure with area $9\times 9$-$\AA^{2}$, solved by
(a) the numerical Poisson solution and (b) the analytical solution. Relevant setups are the same with that in Fig. \ref{poissonsolve1},
except for atom charge $\delta q$.} \label{poissonsolve2}
\end{figure}

Figures \ref{poissonsolve1}-\ref{poissonsolve2} illustrate 2-dimensional
examples with a comparison between the numerically iterative solution and the lumped element model.
In order to connect the spatial-distributive variable $\delta n_{i}\left( \vec{r}\right)$ with the
atom-site notation $\delta q_{i}$, we use
the conventional distribution function for the density function $\delta n_{i}\left( \vec{r}\right)$
in two-dimensional systems:
\begin{equation}
\delta n_{i}\left( r\right) =\frac{ \delta q_{i}}{2 %
\pi \eta^{2}}e^{-\frac{|r-R_{i}|}{\eta}}  \label{disf}
\end{equation}
Here, the circular-symmetry assumption \cite{thesis1} has been adopted,
where $R_{i}$ is the position for atom $i$, and $\eta$ is associated with the effective radius of the MT-sphere by
$\eta\propto r_{MT,i}$ ($\eta= r_{MT,i}$ in this work).
The obtained potential $\delta V_{H}(r)$ is projected on the atomic sites through
\begin{equation}
\delta V_{i}=\frac{\int d\mathbf{r}\delta V_{H}(\mathbf{r})e^{-\frac{%
|r-R_{i}|}{\eta }}}{\int d\mathbf{r}e^{-\frac{|r-R_{i}|}{\eta }}}
\label{projr2a}
\end{equation}
for a comparison with the lumped element model $\mathbf{\vec{V}}$ in this work.
We study two exemplary structures as shown in Figs. (\ref{poissonsolve1}-\ref{poissonsolve2}). The analytical solution presents comparable results with that from the numerically-iterative method.
It is emphasized that the analytical model turns inefficient at large biases or strong density variations,
because the MT sphere cannot accurately account for the distorted and displaced distribution function of the electron density away from the nucleus.

The relevant parameters of the MT radius used in this work are \cite{MTSi,MTO,MTAu,MTP} $r_{MT}(Si)=1.164{\AA}$, $r_{MT}(O)=0.947{\AA}$, $r_{MT}(Au)=1.376{\AA}$, and $r_{MT}(P)=1.377{\AA}$.
All computations are operated on a workstation having 2xCPU(E5-2690 v2) and 128G of DRAM. Fortran source codes can be downloaded online \citep{codeF}.

\section{Time-dependent electron transport in open quantum-dot systems}

\begin{figure}[tbp]
\begin{center}
\includegraphics[scale=0.375]{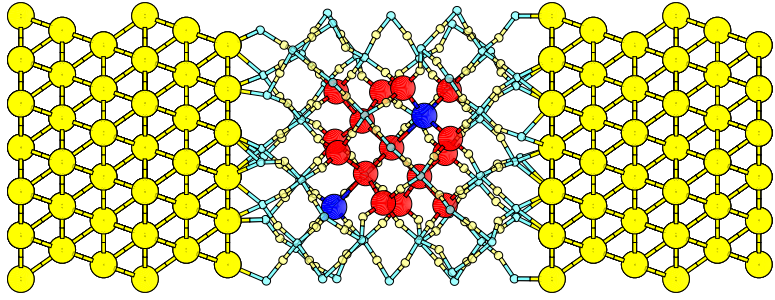}
\end{center}
\caption{Schematics of a Si-based (red atoms) quantum dot embedded in $\beta$-cristobalite SiO$_{2}$ matrix (small light cyan-yellow atoms),
where two dopant atoms (blue phosphorus atoms) are placed inside the quantum dot and at the Si-SiO$_2$ interface, respectively.
The device is enclosed between two semi-infinitely-long Au(111) wires (larger yellow atoms). } \label{fig_qd_dev}
\end{figure}

This section studies the time-dependent electron transport for open quantum-dot systems, as illustrated in Fig. \ref{fig_qd_dev}.
The Si-based quantum dot (red atoms) and SiO$_{2}$ matrix (small light cyan-yellow atoms) in the device region
are enclosed by two semi-infinitely long Au wires. Two dopant atoms (phosphorus; blue atoms) are placed inside the quantum dot
and at the Si-SiO$_2$ interface, respectively, according to their energetically-favored formation energy\cite{QD1}.
It is assumed that the positions of the atoms of Au electrodes are under constraint by the experimental set-ups,
while the atoms of the doped Si-SiO$_{2}$ quantum dot are in equilibrium according to geometry relaxations.
This work initially sets the distance between the nearest cross sections of silica and gold boundaries before geometry relaxations to be 1.8 $\AA$.

The appendix describes in details the other relevant properties, from individual components to the integrated systems.
Additional parameters and numerical techniques are as follows: time step $\delta t=5as$, voltage function $V_{f}(t)=V_{dc}\left[1-exp^{-t/\tau}\right]+V_{ac}sin(\omega t)$
with $\tau=2fs$, the globally-adaptive numerical integral treating Eq. (\ref{a16}), and the fourth-order Runge Kutta methods (RK4)
for solving Eq. (\ref{a3}). Here, we adopt the linear extrapolation of the density matrix $\sigma_{D}$ during the RK4 process.

\subsection{Numerical stability}
\begin{figure}[tbp]
\begin{center}
\includegraphics[scale=0.65]{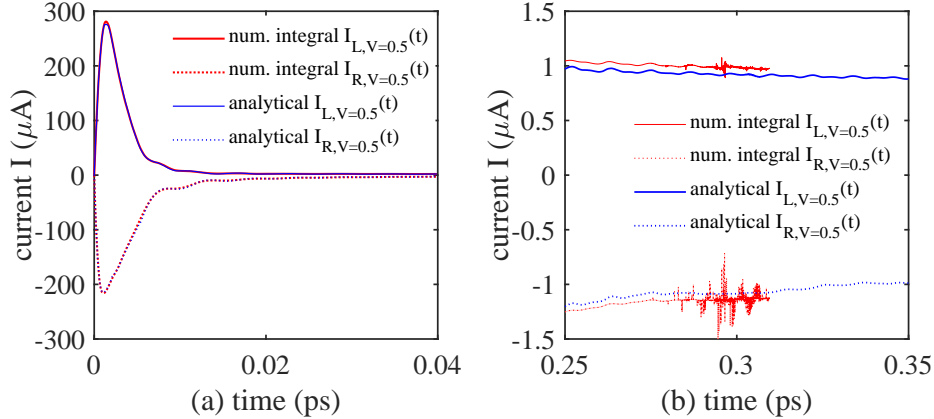}
\end{center}
\caption{Transient current I of the quantum-dot system driven by DC bias $V_{dc}$=0.5V (a) during a finite time period (t$<$0.04 ps) and (b) after a long time period (t$\geq$0.25 ps),
using both a numerical-integral technique on Eq. (\ref{a16})
and an analytical algorithm on Eq. (\ref{add3}).   } \label{fig_T_cmp}
\end{figure}

Figure \ref{fig_T_cmp} shows transient currents of the quantum-dot system driven by bias functions $V_{L}=V_{f,V_{dc}=0.5V,V_{ac}=0V}$ and $V_{R}=V_{f,V_{dc}=-0.5V,V_{ac}=0V}$
for the left and right electrodes, respectively. Calculations by the numerical-integral technique on Eq. (\ref{a16})
and the analytical algorithm on Eq. (\ref{add3}) are compared during a finite time period (t$<$0.04 ps) in Fig. \ref{fig_T_cmp}(a) and after a long time period (t$\geq$0.25 ps)
in Fig. \ref{fig_T_cmp}(b). As indicated in this figure, both methods show transient currents comparable to each other at t$<$0.28 ps.
The calculation by the numerical-integral method, however, begins to abnormally fluctuate after t$\geq$0.28 ps and runs into sudden termination.
The analytical algorithm presents superior numerical stability even at a large time scale, as discussed in the following paragraphs.

\subsection{Transient current driven by DC bias}
\begin{figure}[tbp]
\begin{center}
\includegraphics[scale=0.55]{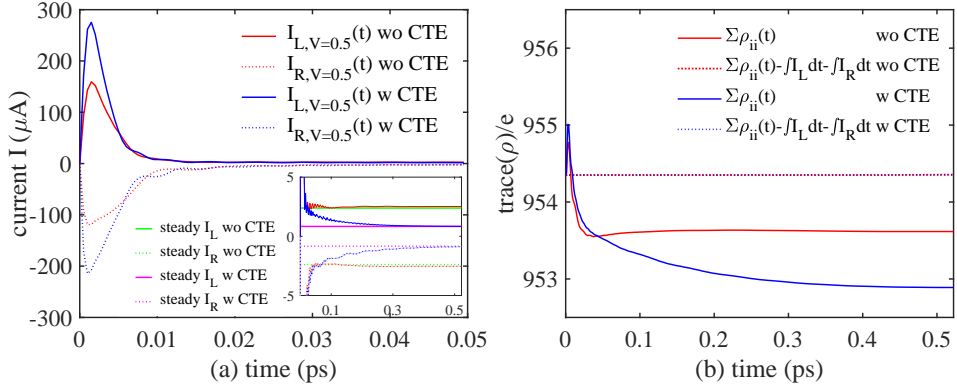}
\end{center}
\caption{(a) Transient current I of the quantum-dot system driven by DC bias $V_{dc}$=0.5V, including and excluding charge transfer effect (CTE).
The inset diagram shows that the currents asymptotically approach the value of the steady solution by SIESTA (green curves).
(b) The corresponding transient charge numbers of the quantum-dot device. The integrals of boundary currents
$\int I_{L}dt+\int I_{R}dt$ are also considered to identify the continuity equation.} \label{fig_TD1}
\end{figure}

Figure \ref{fig_TD1}(a) shows transient currents by analytical TD-NEGF algorithms, both including and excluding corrections of charge transfer effects (CTE).
The voltage functions are set by $V_{L}=V_{f,V_{dc}=0.5V,V_{ac}=0V}$ and $V_{R}=V_{f,V_{dc}=-0.5V,V_{ac}=0V}$ for the left and right electrodes, respectively.
For comparison and validation, we calculate the corresponding steady currents with the Landauer Buttiker formula \cite{datta1}, an integral of the transmission functions in Fig. \ref{fig_T}(b), using the SIESTA program.
With the steady and transient results in Figure \ref{fig_TD1}(a),
one observes that the transient currents asymptotically approach the values of the corresponding steady solutions \cite{longtime1,longtime2}
(see the inset diagram), no matter whether or not the charge transfer effects are synchronously considered. In fact, the inclusion of charge transfer effects presents considerable corrections
for the convergence of the transient current, suggesting non-trivial influences of charges beyond the ground state approximation. The curves also depict that
the calculation including CTE requires a much longer time to bring the system into the steady sate, inferring a self-consistent redistribution process of the device charge.
Figure \ref{fig_TD1}(b) shows the transient properties of the electron number and the integrals of boundary currents, obeying the continuity equation for the device region.

\subsection{Transient current driven by AC bias}
\begin{figure}[tbp]
\begin{center}
\includegraphics[scale=0.5]{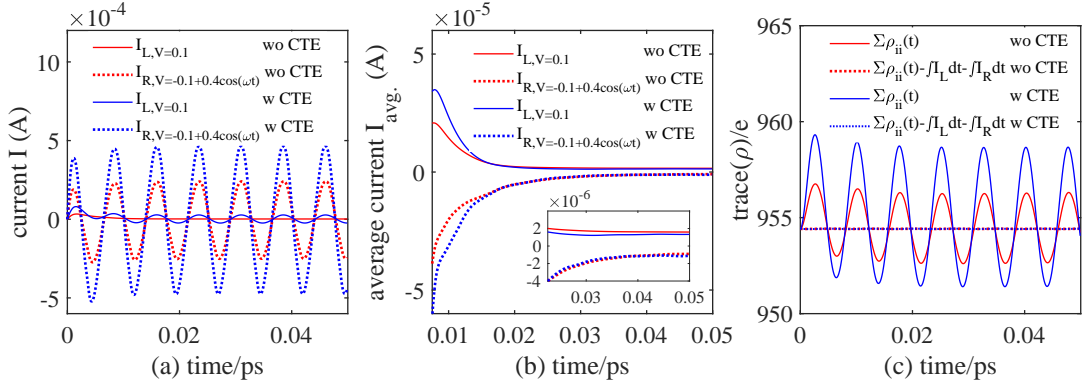}
\end{center}
\caption{(a) Transient current I of quantum-dot devices driven by bias $V_{L}$=0.1V and $V_{R}$=-0.1+0.4$sin(\omega t)$ V, both
including and excluding the charge transfer effect (CTE).
(b) The net currents $I_{avg.}$ by averaging $I$(t) of (a) over period $T=2\pi/\omega$. The inset diagram shows the asymptotical net currents.
(c) The corresponding transient charge numbers of the quantum-dot device. The integrals of boundary currents
$\int I_{L}dt+\int I_{R}dt$ are also considered to identify the continuity equation.} \label{fig_TD2}
\end{figure}

Figure \ref{fig_TD2} studies the transient currents for the quantum-dot devices driven by AC voltages.
To observe the properties of charges inside the device, the voltage functions are asymmetrically set
by $V_{L}=V_{f,V_{dc}=0.1V,V_{ac}=0V}$ and $V_{R}=V_{f,V_{dc}=-0.1V,V_{ac}=0.4V}$ for the left and right electrodes, respectively.
The AC frequency is $\omega=0.8\times 10^{15}$ Hz. It is noted that the AC signaling is only applied on the right electrode.
In Fig. \ref{fig_TD2}(a), the calculation including charge transfer effects exhibits an oscillating interfacial current $I_{L}$ and represents the physical AC signaling through
the device. The calculation excluding charge transfer effects, however, depicts a constant interfacial current $I_{L}$, exhibiting non-physical blocking of AC signals.
Figure \ref{fig_TD2}(b) studies the net current $I_{avg.}(t)$ by averaging $I(t)$ of Fig. \ref{fig_TD2}(a) over one period $T=2\pi/\omega$.
On the basis of the analyses above, we conclude that the charge transfer effects considerably influence the transient properties of the devices,
but its significance on the steady outcomes
remains unrevealed. Relevant discussions by means of photon-assisted dynamics will be discussed in the following paragraph.
Here, similar to the analysis on the DC condition, Figure \ref{fig_TD2}(c) monitors the validity of the algorithms by the continuity equation.

\subsection{Photon-assisted transport}
\begin{figure}[tbp]
\begin{center}
\includegraphics[scale=0.6]{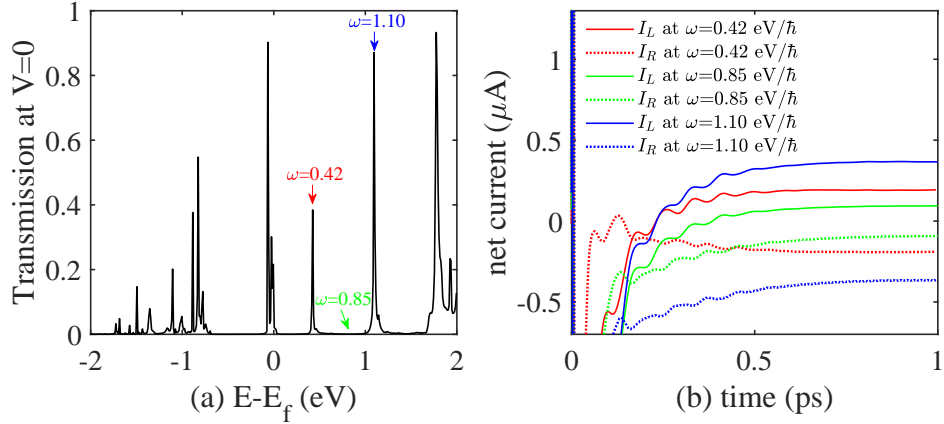}
\end{center}
\caption{(a) Transmission function of the open quantum-dot system at V=0.0 eV. Three energy levels representing states inside the first excited energy-band($\hbar\omega=0.42eV$),
inside the energy gap($\hbar\omega=0.85eV$) ,
and inside the second excited energy-band($\hbar\omega=1.10eV$), respectively, are indicated in the diagram.
(b) The transient net currents driven by biases $V_{L}$=$V_{dc}+V_{ac}sin(\omega t)$ and $V_{R}$=$-V_{dc}+V_{ac}sin(\omega t)$ with frequencies given in (a).} \label{fig_PAT}
\end{figure}
Figure \ref{fig_PAT} studies the photon-assisted transport (PAT) of the quantum-dot devices by applying AC voltages at specified frequencies.
The voltage functions are set
by  $V_{L}=V_{f,V_{dc}=0.035V,V_{ac}}$ and $V_{R}=V_{f,V_{dc}=-0.035V,V_{ac}}$, with condition $eV_{ac}=\hbar\omega\cdot sin(\omega t)$.
To identify proper AC frequencies for photon excitations, Fig. \ref{fig_PAT}(a) calculates the zero-bias transmission function.
Here, we select three energy levels as $\hbar\omega=0.42eV$, $\hbar\omega=0.85eV$, and $\hbar\omega=1.10eV$, representing the states inside the first excited energy-band,
inside the energy gap, and inside the second excited energy-band, respectively. Figure \ref{fig_PAT}(b) displays the transient net currents driven by the voltage function with given frequencies.
Numerical results indicate that the photons with energies meeting excited levels ($\hbar\omega=0.42eV$ and $\hbar\omega=1.10eV$)
can distinctly enhance electron transport and raise the net DC current;
otherwise, the photon ($\hbar\omega=0.85eV$) presents less significant influences on the current. Another exemplary device of
4,4'-Bipyridine molecules is addressed in the appendix for more detailed discussions about PAT with the Tien-Gordon approach.

\section{Conclusions}
This research presents analytical algorithms, with fortran codes, to study excitation transports in quantum devices.
Relevant analyses show that the algorithms enable efficient and numerically-stable computations even at large time and space scales, whereas
conventional treatments could suffer problems on numerical divergence and high-demanding computation cost.
We also consider the effects of transiently-transferring charges, inferring to excitations or populations of electrons beyond ground states,
together with a lumped element model. The validity of this work is discussed with a comparison with time-independent density functional calculations
and the photon-assisted transport dynamics.

\section{Acknowledgement}
This work was supported by ChiMei Visual Technology Corporation under Project no. 37.

\appendix
\section{Physical properties from individual components to integrated quantum systems}
\subsection{Atomic Electrodes: Au(111) Nanotubes}
\begin{figure}[tbp]
\begin{center}
\includegraphics[scale=0.42]{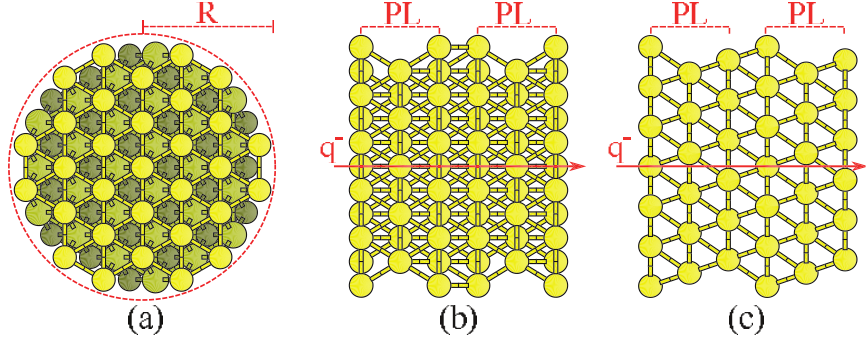}
\end{center}
\caption{Ball-stick representation of the Au(111) nanotube (a) in the longitudinal perspective and (b-c) in two lateral perspectives.
The radius of the cross section in (a) is set as R=2a.
The red arrow signifies the quasi one dimensional (red arrow) charge transport.} \label{fig_au111}
\end{figure}
\begin{figure}[tbp]
\begin{center}
\includegraphics[scale=0.6]{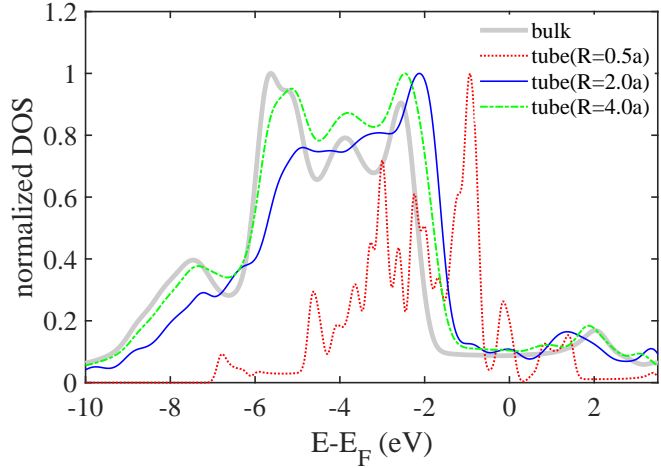}
\end{center}
\caption{Normalized density of states (DOS) for Au bulk and infinite Au(111) nanotubes, in which the radiuses of the nanotubes are set as R=0.5a, R=2.0a, and R=4.0a.} \label{fig_DOSau111}
\end{figure}

This research uses Au(111) nanotubes as atomic electrodes. The length $\ell$ of the Au-Au bond is determined with geometry relaxations of the Au bulk
in the SIESTA program\cite{siesta1,siesta2}, obtaining the value $\ell$=2.8785 ${\AA}$ (lattice constant a=$\sqrt{2}\ell$=4.0708 ${\AA}$, which is similar to the experimental value \cite{chembook1} of 4.0782 ${\AA}$). The effects of core electrons are evaluated with norm-conserving pseudopotentials in the local density approximation (Ceperley-Alder exchange-correlation potential\cite{LDA1,LDA2}), which are generated by the ATOM program\cite{atom1,siesta1}. The valence electrons of Au are calculated in the s-d hybridized configuration \cite{sdhybrid1}. All the calculations for nanotubes are performed on $8\times8\times8$ Monkhorst-Pack grids in reciprocal spaces under an electronic temperature of 300K. Figure \ref{fig_au111} shows (a) the longitudinal perspective and (b-c) two lateral perspectives for a finite segment of Au(111) nanotubes. In actual computations, the nanotube is set as an infinite stack of principle layers (PL) along the axial (longitudinal) direction, and has cross-section radius R. Figure \ref{fig_DOSau111} shows the normalized density of states (DOS) for Au bulk and Au(111) nanotubes, where the radiuses of the nanotubes are set as R=0.5a, R=2.0a, and R=4.0a, respectively.
Here, $E_{F}$ is the Fermi level corresponding to the mentioned system. In Fig. \ref{fig_DOSau111}, DOS of the Au bulk shows metallic properties as the literature \cite{AuBulkDOS} reports. For Au(111) nanotubes, when increasing the cross-section radius R, the DOS functions of the tubes at energies near $E_{F}$ change from discrete to uniform distributions, depicting the transfer of systems from 1D-line to 3D-bulk structures. In this work, we use Au(111) nanotubes with R=2a for semi-infinite electrodes in transport problems. This adoption (setting R=2a) meets the requirement of slowly-varying DOS for the wide-band limit (WBL) condition \cite{WBL1}, and demands computation resources that are affordable.

\subsection{Doped Si-SiO$_{2}$ quantum dots}

\begin{figure}[tbp]
\begin{center}
\includegraphics[scale=0.25]{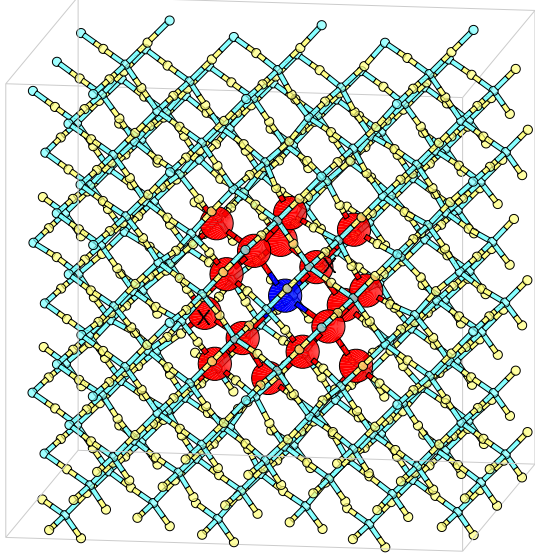}
\end{center}
\caption{Schematics of Si Quantum Dot (red atoms) embedded in SiO$_{2}$ matrix (light cyan-yellow atoms). The phosphorus atom P (blue atom) is doped inside the quantum dot for the 1P-doping condition. Two red atoms with X marks denote the doping locations inside the quantum dot and at the interface, respectively, for the 2P-doping condition.} \label{fig_cell}
\end{figure}

\begin{figure}[tbp]
\begin{center}
\includegraphics[scale=0.8]{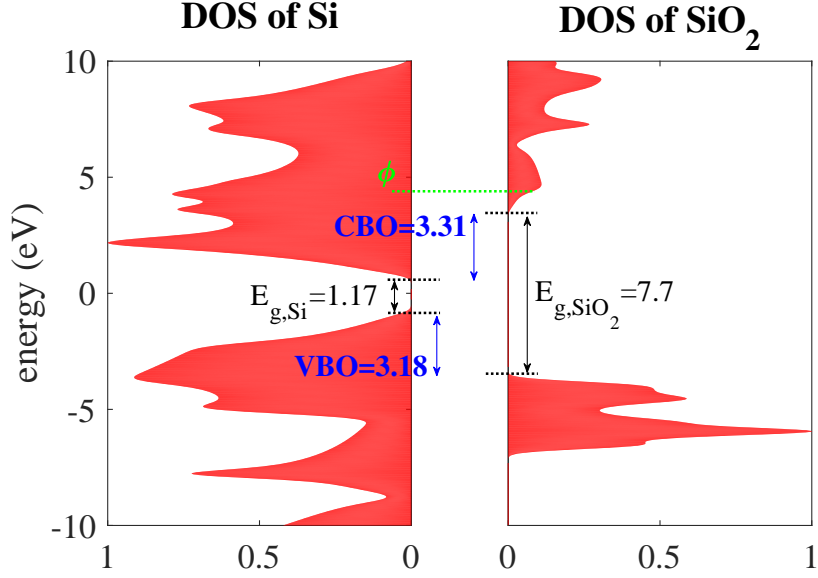}
\end{center}
\caption{Band diagrams of Si-SiO$_{2}$-slabs heterojunction by Anderson's rule. The density of state at equilibrium is arranged according to a hypothetical flat vacuum level.
The computed energy gaps are $E_{g,Si}$=1.17 eV and $E_{g,SiO_{2}}$=7.7 eV. The valence band offset VBO is 3.18 eV and the conduction band offset CBO is 3.31 eV.} \label{fig_DOSsisio2}
\end{figure}

This research investigates the silicon quantum dots with diameters around 1.0 nm that are embedded in a $\beta$-cristobalite SiO$_{2}$ matrix.
The dopant phosphorus (P) atoms are placed inside quantum dots based on their energetically-favored formation of structures \cite{QD1} (see Fig. \ref{fig_cell}).
Lattice constants are determined with geometry relaxations in the SIESTA program (setting orbital bases s and p for species Si, O, and P).
The obtained values are 5.5001 $\AA$ (5.4306 by experiment \cite{chembook1})
for the Si diamond structure and 7.46831 $\AA$ (7.160-7.403 $\AA$ in textbooks \cite{chembook1,sio2a}) for  $\beta$-cristobalite silica.
We investigate the energy band diagram of Si-SiO$_2$-slabs heterojunctions by using Anderson's rule through Fig. \ref{fig_DOSsisio2},
in which the vacuum levels (green dotted lines) of Si and SiO$_2$ slabs are aligned at the same energy. Here, the vacuum level is defined
as the effective potential $\phi$ (adding local pseudopotential, Hartree potential, and exchange-correlation potential) at zero-density points near
the surface of slabs having 35 atomic layers.  All calculations are performed at $\Gamma$-point of the reciprocal space. As indicated in Fig. \ref{fig_DOSsisio2},
the vacuum levels are 1.064 eV and 1.626 eV for Si-slab and SiO$_{2}$-slab, respectively, corresponding to working functions $W_{Si}$=4.46 eV and $W_{SiO_{2}}$=4.52 eV.
The experimental value \cite{chembook1} is $4.60\leq W_{Si}\leq 4.91$ eV. The computed energy gaps are 1.17 eV for bulk silicon and 7.7 eV for $\beta$-cristobalite silica,
which can be compared to the experimental values of 1.1 eV and 9.0 eV, respectively.
The valence band offset (VBO) and conduction band offset (CBO) for Si-SiO$_{2}$ heterojunctions are estimated to be 3.18 eV and 3.31 eV, respectively.
The obtained VBO values are smaller than experimental measurements \cite{vbo1,vbo2} with VBO=4.6 eV and CBO=3.1 eV. Several theoretical works
using hopping mechanisms \cite{QD1,QD2,QD3} give VBO$\approx$2.6 eV and CBO$\approx$3.9 eV.

\begin{figure}[tbp]
\begin{center}
\includegraphics[scale=0.6]{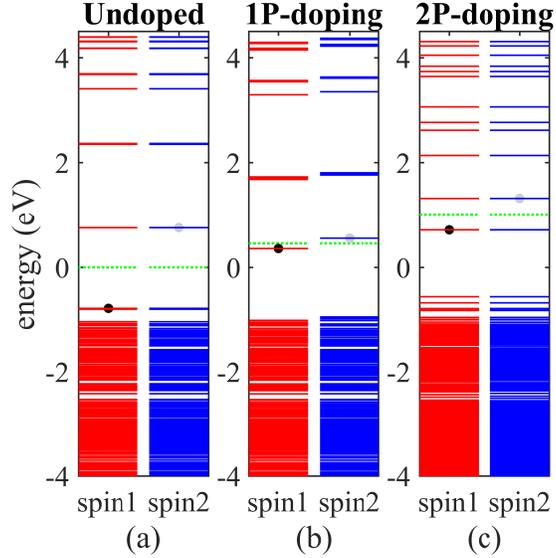}
\end{center}
\caption{Spin-up and spin-down spectra of (a) undoped, (b) 1P-doping, and (c) 2P-doping systems. Energies are aligned using the embedding SiO$_2$ states, and are shifted with the reference of the fermi level of the undoped structure. Black and gray circles mark HOMO and LUMO states, respectively. The green dotted line represents the fermi level of the corresponding system.} \label{fig_eigval}
\end{figure}

With relevant material parameters, the Si-SiO$_{2}$ quantum-dot device in Fig. \ref{fig_cell} is constructed from a $3\times3\times3$ supercell of $\beta$-cristobalite silica
by removing O atoms in a cut-off box \cite{QD1}. Figure \ref{fig_eigval} reports the eigenvalue spectra for the undoped, 1P-doping, and 2P-doping structures after relaxation processes,
using the corresponding initial geometries in Fig. \ref{fig_cell}. The spectrum energies are aligned along the level of the deep valence states of SiO$_2$,
and the origin of the energy axis is determined according to the fermi level of the undoped structure. Black and gray circles mark the highest occupied molecular orbital
(HOMO) and lowest unoccupied molecular orbital (LUMO) states, respectively. The green dotted line represents the fermi level of the corresponding structure.

In Fig. \ref{fig_eigval}(a), the undoped quantum-dot structure exhibits a distinguished energy spectrum from that of the slab-heterojunction in Fig. \ref{fig_DOSsisio2},
revealing the interfacially strain-related electron levels \cite{mismatch1}. For the 1P-doping system, the odd number of electrons leads to the spin-dependent energy spectrum
in Fig. \ref{fig_eigval}(b), which depicts a clear donor behavior and agrees well with previous works \cite{QD1,donor1}. This study adopts the 2P-dopping structure in
Fig. \ref{fig_eigval}(c) due to the following considerations: (i) has lower threshold voltage owing to the rising fermi level and the decreasing energy gap,
compared to the other two structures; and (ii) has spin independence for reduced dimensions of the atomic orbital sets and the negligible spin-flip mechanism.

\subsection{Transmission function of the open quantum-dot system}
\begin{figure}[tbp]
\begin{center}
\includegraphics[scale=0.55]{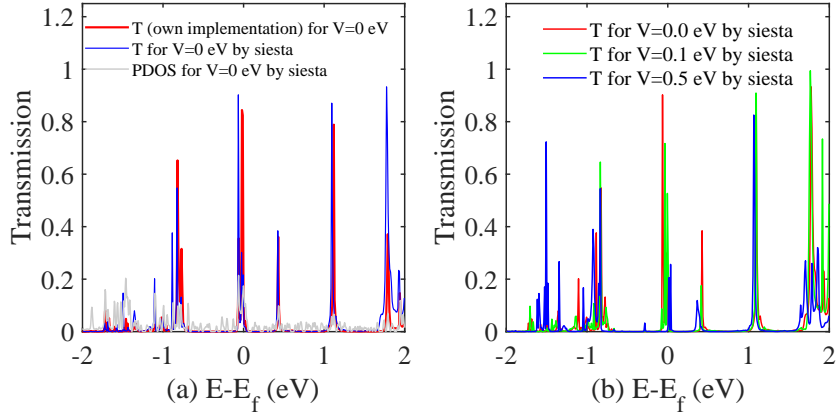}
\end{center}
\caption{(a) Transmission functions of the quantum-dot system (V=0.0 eV) using SIESTA and the fortran program, to be compared with the projected density of state (PDOS) of Si-SiO$_{2}$ QDs. (b) Transmission functions of the quantum-dot system calculated by SIESTA programs at different voltage biases.  } \label{fig_T}
\end{figure}

The complete open quantum-dot system is depicted in Fig. \ref{fig_qd_dev}.
Its transmission function T (blue curve) is calculated by SIESTA::Transiesta programs, and is compared with
the projected density of the state (PDOS; gray curve) of the Si-SiO$_2$ quantum dot, as shown in Fig. \ref{fig_T}(a). The red-curve is calculated by fortran program
using the tight-binding formulation.
In Fig. \ref{fig_T}(b), the transmission functions for the system with different biases are computed by SIESTA,
signifying the effects of non-conserved charges in open quantum systems.

\section{Photon-assisted transport in 4,4'-Bipyridine molecules}
\begin{figure}[tbp]
\begin{center}
\includegraphics[scale=0.27]{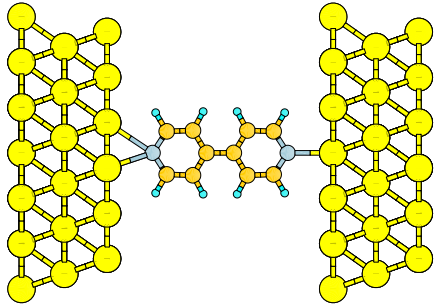}
\end{center}
\caption{Schematics of the 4,4'-Bipyridine molecule device (carbon=orange atoms; hydrogen=cyan atoms; nitrogen=gray atoms).
The device is enclosed between two semi-infinitely-long Au(111) wires (yellow atoms) with given applied voltages.} \label{fig_Bi1}
\end{figure}

This appendix discusses photon assisted transport in the molecule device, as illustrated in Fig. \ref{fig_Bi1}.
The 4,4'-Bipyridine molecule in the device region
is enclosed by two semi-infinitely long Au wires.
It is assumed that the positions of the atoms of Au electrodes are under constraint by the experimental set-ups,
while the device atoms are in equilibrium according to geometry relaxations.
The distance between the nearest device atoms and gold boundaries is initially set to be 2.5 $\AA$ for setting weak device-electrode couplings.
Parameters about the muffin-tin radius are referred to the literature \cite{mt_radius}.

\begin{figure}[tbp]
\begin{center}
\includegraphics[scale=0.55]{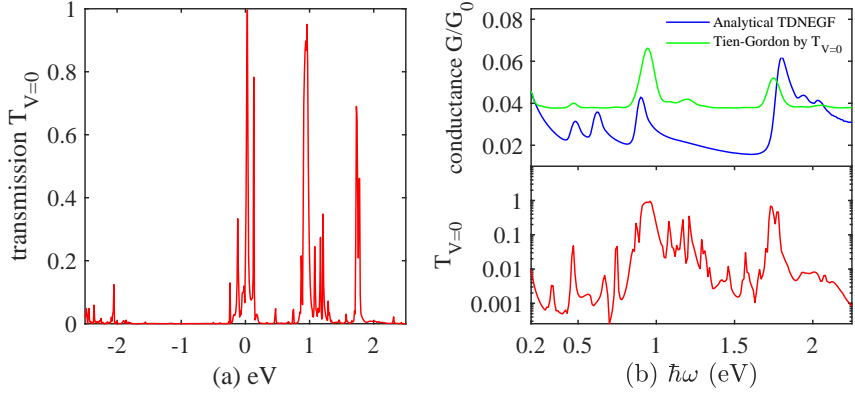}
\end{center}
\caption{(a) The zero-bias transmission function of the 4,4'-Bipyridine molecule device.
(b) The upper diagram calculates the device conductance $G$ as a function of the photon energy $\hbar\omega$ by Tien-Gordon approximation and analytical TD-NEGF algorithms.
The log-scale transmission curve in the lower diagram is shown for comparison.
} \label{fig_Bi2}
\end{figure}

To analyze the device conductance with respect to photon energies, we set the voltage functions
by $V_{L}=V_{f,V_{dc}=0.05V,V_{ac}}$ and $V_{R}=V_{f,V_{dc}=-0.05V,V_{ac}}$, with condition $eV_{ac}=\hbar\omega\cdot sin(\omega t)$.
Figure \ref{fig_Bi2}(a) shows the complete zero-bias transmission function of the system, depicting accessible transport channels in the molecule device.
In Fig. \ref{fig_Bi2}(b), the upper diagram calculates the device conductance (blue curve) as a function of the photon energy using the TD-NEGF algorithm,
and compares that (green curve) by the Tien-Gordon approach. Here, $G_{0}=2e^{2}/h=7.748\times 10^{-5}S$ denotes the conductance quantum.
Numerical results demonstrate that both curves show quantitative agreements in the
low frequency regime. Beyond the linear response (low frequency) condition by the Tien-Gordon approximation, the conductance functions present qualitative comparability only with
photon energies near primary excited energy-bands. The lower diagram in Fig. \ref{fig_Bi2}(b) plots the corresponding log-scale transmission curve,
which is similar to that via TD-NEGF algorithms.

\end{document}